**Flexible water excitation for fat-free MRI at 3 Tesla using lipid insensitive binomial off-resonant RF excitation (LIBRE) pulses**

Bastiaansen et al: LIBRE for fat free MRI

Jessica AM Bastiaansen[1*], Matthias Stuber[1,2]

[1] Department of Radiology, University Hospital (CHUV) and University of Lausanne (UNIL), Lausanne, Switzerland

[2] Center for Biomedical Imaging, Lausanne, Switzerland

[*]Correspondence: Jessica AM Bastiaansen, Department of Radiology, University Hospital Lausanne (CHUV), Rue de Bugnon 46, BH 8.84, 1011 Lausanne, Switzerland, Phone: +41-21-3147516, Email: jbastiaansen.mri@gmail.com

Word count manuscript (body text): 5141

Figure count: 8

Table count: 2

Supporting figure count: 2

# ABSTRACT

**Purpose**: To develop a robust and flexible low power water excitation pulse that enables effective fat suppression at high magnetic field strength.

**Methods**: A water excitation method that uses spatially non-selective pulses was optimized in numerical simulations, and implemented and tested in phantoms and healthy volunteers at 3T. The lipid insensitive binomial off-resonant excitation (LIBRE) pulse comprises two low power rectangular sub-pulses that have a variable frequency offset, phase offset and duration. The capability and extent of LIBRE fat suppression was quantitatively compared with conventional fat saturation (FS) and water excitation (WE) techniques.

**Results**: LIBRE enables simultaneous water excitation and near complete fat suppression in large volumes at 3T as demonstrated by numerical simulations, and experiments. In phantoms and in human subjects, the frequency responses matched well with those from the numerical simulation. Comparing FS and WE, LIBRE demonstrated an improved robustness to magnetic field inhomogeneities, and a much more effectively suppressed fat signal. This applied for a range of pulse durations and pulses as short as 1.4 ms.

**Conclusion**: A flexible water excitation method was developed that shows robust, near complete fat suppression at 3T.

## INTRODUCTION

The suppression of fat signals is a critical component of many MRI exams in the clinic. However, and despite the availability of many elegant approaches, fat signal suppression remains often challenging – especially at high magnetic field strength and in regions of increased field inhomogeneities.

Techniques aimed at suppressing fat mainly take advantage of the two characteristic differences in the behavior of water and fat (1). 1) Either they make use of the difference in relaxation times or 2), they exploit chemical shift, 3) or both. Fat suppression may consist of the application of pre-pulses that result in nulling of the fat signals during MR image data collection, such as chemical shift selective (CHESS) radio frequency (RF) saturation pulses (2-5) or inversion recovery (IR) pulses (6-8). Some approaches use the difference in magnetization evolution between consecutive acquisitions such as Dixon-based techniques (9-11) or IDEAL (12,13). Although these methods attempt to separate the water and fat signals to quantify their proportions, they do not necessarily aim for fat suppression. Other methods use specific radiofrequency (RF) excitation pulses that are water selective. This includes binomial RF excitation pulse schemes that leave the fat magnetization unperturbed while exciting the water magnetization only (14). On the one hand, such water selective excitation (WE) affords the advantage that magnetization preparation can be avoided and a more time efficient data collection may be obtained. On the other hand, prolonged RF pulse durations are inevitable and the resulting increase in TR has to be considered. WE may consist of rectangular pulses that also have the potential to suppress specific signals by using their off-resonance frequency response (15,16). Conventional WE approaches using binomial pulse pairs often show narrow bandwidths for fat suppression and are relatively sensitive to the main magnetic field ($B_0$) inhomogeneities, since they rely on an exact phase evolution between the water and fat resonances during the interpulse delay time (14,17-19). The simplest binomial WE pulse consists of a 1-1 pulse pair with an interpulse delay time allowing a 180° phase evolution of water and fat. The interpulse delay time can be shortened by applying a phase-modulation of the different RF sub-pulses, although at the expense of the effective RF excitation angle (19). Methods that utilize short binomial pulses without interpulse delay have been demonstrated to allow very rapid spectrally selective excitation, but also at the expense of rather narrow fat suppression bandwidths (20). Recently BORR was proposed, a method in which two consecutive adjacent rectangular pulses with a 180° phase shift are applied (21). This method shows large suppression bandwidths for fat signals in the numerical simulations, and was applied in the human breast and eye at 3 T (21,22), but relatively lengthy RF pulse durations of 2.6 ms and 3.2 ms were used.

Unfortunately, many of the above methods for fat suppression remain sensitive to magnetic field inhomogeneities, RF pulse imperfections and large RF energy deposits. And this becomes particularly challenging at higher magnetic field strength. In addition, the inversion recovery time needed for $T_1$ based fat signal-nulling may lead to time-inefficient data collection, and sometimes recovery of the fat magnetization during prolonged acquisition windows leads to ineffective fat signal suppression.

The aim of this study was therefore the development of a new RF water excitation pulse, which selectively suppresses fat and excites water, which is robust to magnetic field inhomogeneities, which supports short RF pulse durations, which remains effective at high magnetic field strength, and which does not add an extra burden in terms of the specific absorption rate. This pulse was developed using numerical simulations to analyze the response of fat and water resonances and to help guide the pulse parameter settings. This was followed by in vitro experiments at 3 T and then by first in vivo human studies. In these experiments, we thoroughly characterized the proposed excitation scheme and tested the hypothesis that these Lipid Insensitive Binomial Off-Resonance Excitation (LIBRE) pulses remain effective over both a relatively broad range of $B_0$ and $B_1$ as well as pulse durations.

## METHODS

### *LIBRE pulse implementation*

The design of this new RF water excitation pulse consists of a pair of frequency and phase modulated rectangular RF excitation pulses without inter pulse delay time. The two pulses have an identical RF excitation angle α, frequency offset f, yet different phases $\varphi_1$ and $\varphi_2$. The relative phase shift of the second sub-pulse ($\varphi_2$) with respect to the phase of the first sub-pulse ($\varphi_1$) is determined by the frequency offset f, and duration $\tau_1$ of the first pulse:

$$\varphi_2 = \varphi_1 + 2\pi f \tau_1 \qquad\qquad (Eq.\ 1).$$

During the first RF sub-pulse, which is off-resonant, a rotation of the tissue magnetization will occur. By applying a RF phase offset that equals $\varphi_2$ on the second RF sub-pulse one may maximize the water excitation. The entire rotation during the LIBRE pulse is described by Eq. 6 and is not intuitive. Therefore, numerical simulations were performed to clearly illustrate this concept and are described in a later section. Here, the use of $\varphi_1$ and $\varphi_2$ describes a relative phase change between the two sub-pulses so that the condition may take into account RF phase cycling.

### *Background*

For a rectangular pulse, the RF excitation angle α is proportional to the field strength $B_1$, the duration of the applied RF pulse t, and the gyromagnetic ratio γ. Therefore, $\alpha = 2\pi\gamma B_1 t$, with α expressed in radians.

In this particular situation, off-resonance effects cannot be neglected and are even exploited in the LIBRE implementation to null the fat resonances. In a rotating frame of reference, let $\Delta f$ be the difference between the RF excitation pulse and resonance frequency of the tissue, then the magnetization will rotate about an effective magnetic field $B_e$, with:

$$B_e(t) = \begin{pmatrix} B_1(t)\cos(\varphi) \\ B_1(t)\sin(\varphi) \\ \Delta f \,/\, \gamma + \Delta B_0 \end{pmatrix} \qquad \text{(Eq. 2)},$$

With $\varphi$ equal to $\varphi_1$ during the first sub-pulse and equal to $\varphi_2$ (Eq. 1) during the second sub-pulse. The magnitude of the effective field $B_e$, is then described by:

$$\left| B_e(t) \right| = \sqrt{ B_1(t)^2 + \left( \frac{\Delta f}{\gamma} + \Delta B_0 \right)^2 } \qquad \text{(Eq. 3)},$$

where $\Delta B_0$ expresses the local magnetic field inhomogeniety. Taking into account these off-resonance effects, the frequency at which the magnetization precesses around the effective field is $\gamma B_e$, and the effective angular rotation angle $\psi_e$ about $B_e$ during each rectangular sub-pulse with duration $\tau$ is described by:

$$\psi_e(\tau) = \int_0^\tau 2\pi\gamma B_e(t)dt = 2\pi\gamma \left| B_e \right| \tau \qquad \text{(Eq. 4)}.$$

Intuitively, one can consider lipid magnetization undergoing a full rotation ($\psi_e(\tau)=2\pi$) around $B_e$ to not have a transverse component at all and thus leading to fat signal suppression. Taking into account a single rectangular sub-pulse that subjects the magnetization to a single rotation ($2\pi$), the condition for optimal fat suppression can be obtained after rewriting (Eq. 4) and is similar to that described in (16), resulting in the following relation between RF offset frequency, nominal excitation angle $\alpha$ and duration for fat suppression of a single sub-pulse:

$$\tau = \sqrt{1-(\alpha\,/\,2\pi)^2} \,/\,(\Delta f + \gamma\Delta B_0) \qquad \text{(Eq. 5)}.$$

However, a complete description of the actual rotation about the off-resonance field $B_e$ during the two sub-pulses is not intuitive and  was obtained by the use of rotation matrices (23). The evolution of the tissue magnetization $M_{tissue}$ following the entire LIBRE pulse can be described by

$$M_{tissue}(\tau_1+\tau_2) = R_Z(\varphi_2)R_{B_e}(\psi_e(\tau_2))R_Z(-\varphi_2)R_{B_e}(\psi_e(\tau_1))M_0 \qquad \text{(Eq. 6)},$$

where $R_{Be}$ describes the rotation about the effective magnetic field $B_e$, $R_Z$ describes the rotation about the z-axis as the second sub-pulse has a phase offset $\varphi_2$, and $M_0$ is the thermal equilibrium magnetization of the respective tissue.

*Numerical simulations*

Simulations were performed in Matlab (The MathWorks, Inc., Natick, MA, United States) to predict and characterize the magnetization behavior of both water and fat components, following water excitation using the LIBRE pulses. These simulations instruct about the range of optimal parameters settings on the MR system, as well as characterize the response as a function of magnetic field inhomogenieties. The transverse and longitudinal magnetization components, as well as the phase (Supporting Fig. S1), of water and fat were calculated and evaluated with (Eq. 6) as a function of the pulse offset frequency f, duration $\tau$, magnetic field inhomogeneities $\Delta B_0$, and RF excitation angle $\alpha$. The simulation of the RF excitation angle $\alpha$ will further provide information about the sensitivity of the LIBRE pulse to $B_1$ inhomogeneities.

Simulation of the water and fat magnetization evolution was performed and expanded to take into account longitudinal ($T_1$) and transverse ($T_2$) relaxation times during repeated excitations using the Bloch equations (24). The $T_1$ of muscle was assumed to be 1500 ms, $T_2$ 50 ms (25), repetition time 6 ms, and number of RF excitations 50. In the numerical simulation, the resonance frequencies of water and of fat at 3 T were set to 0 Hz and -440 Hz respectively. This simulation was performed for RF excitation angles $\alpha$ ranging from 0.5° to 40°, $\gamma\Delta B_0$ from -300 to 300 Hz, RF pulse durations $\tau$ from 0.1 to 2 ms, and RF pulse offset frequencies f from -1000 to 1000 Hz. Perfect spoiling of transverse magnetization was assumed by setting the transverse magnetization to zero after each excitation. Then plots were made to visualize the transverse magnetization response ($M_{XY,water}$, $M_{XY,fat}$, $M_{XY,water}$-$M_{XY,fat}$) of 1) varying $\alpha$ (with fixed $\gamma\Delta B_0$=0 and $\tau$=1.1 ms ), 2) varying $\tau$ (with fixed $\gamma\Delta B_0$=0 and $\alpha$=10°), and 3) varying $\gamma\Delta B_0$ (with fixed $\alpha$=10° and $\tau$=1.1 ms) as function of the relative pulse offset frequency f. The resultant plots describe the behavior of the LIBRE pulse as a function of these parameter settings. Since the transverse magnetization correlates with the measured signal during MRI acquisitions, these plots help illustrate the optimal range of LIBRE pulse parameters that can be used for in vitro and in vivo experiments. For further comparison, numerical simulations were also performed for the BORR pulse (21) using the same scan parameters. In these simulations the relative phase shift of the second sub-pulse ($\varphi_2$) with respect to the first sub-pulse ($\varphi_1$) was fixed to $\pi$ as described in (21).

To illustrate the concept that water excitation is maximized using a LIBRE defined phase offset (defined by Eq. 1), simulations were performed to calculate the tissue magnetization after RF excitation with two off-resonant sub-pulses where the relative phase offset of the second sub-pulse $\varphi_2$ was varied over a range from 0 to 2π (360°) in respect to the first. $M_{xy}$ was then plotted as function of the tissue frequency and the phase offset of the second sub-pulse ($\varphi_2$). A selection of four different combinations of pulse parameters was studied that, according to the previous simulations, showed optimal fat suppression: RF pulses with a sub-pulse duration ($\tau$) and RF frequency offset (f) of 1.3 ms and 300 Hz, 1.1 ms and 500 Hz, 0.9 ms and 700 Hz, and 0.7 ms and 1000 Hz, respectively. The simulation was performed for a single RF excitation with α=10°.

*MRI phantom study*

The LIBRE pulses were implemented as part of a Cartesian 3D segmented k-space spoiled gradient recalled echo (GRE) MRI sequence (Fig. 1). To this end, the original excitation pulse was replaced with the LIBRE pulse and installed on a 3.0 T clinical MRI scanner (MAGNETOM Prisma, Siemens AG, Erlangen, Germany). LIBRE pulse parameters such as the duration, RF excitation angle, inter-pulse spacing, and offset frequency are adjustable on the user interface of the scanner.

A selection of four LIBRE pulse parameters was evaluated that, according to the numerical simulations, showed optimal fat suppression (Fig. 2). To this end, experiments were performed with a LIBRE pulse duration (2$\tau$) of 2.6 ms and a 300 Hz frequency offset, 2.2 ms and 500 Hz, 1.8 ms and 700 Hz, and 1.4 ms and 900 Hz, respectively (indicated by stars in Fig. 2). For these four pulses, the relative phase shift of the second pulse ($\varphi_2$) with respect to the first pulse ($\varphi_1$) was 1.4π, 1.3π, 1.1π, and 0.8π respectively. These MRI experiments were performed on a three-compartment cylindrical phantom containing mixed solutions of agar, $NiCl_2$ (Sigma Aldrich, St. Louis, MO), and baby oil (Johnson and Johnson, New Brunswick, NJ)(26). These compartments mimic the magnetic relaxation properties ($T_1$, $T_2$) of muscle, blood, and fat. The phantom measurements were repeated five times, with a field of view (FOV) of $122 \times 176 \times 160$ mm³, a matrix size of $122 \times 176 \times 160$ (resulting in a voxel size of 1.0 x 1.0 x 1.0 mm³), 50 k-space lines per segment using a linear reordering approach, RF excitation angle = 10°, using RF and gradient spoiling.

Further experiments were performed to compare the LIBRE method with 1) a standard MRI acquisition in which no fat suppression technique is applied, 2) a conventional fat suppression (FS) method that uses a CHESS pulse (4) to null the fat resonances prior to the imaging sequence, and 3) a conventional water excitation (WE) method that uses a 1-1 binomial RF pulse pattern. Scan parameters are summarized in Table 1.

For each MRI acquisition, the shortest possible TR was used, which required the addition of a dead time each 50 TRs in some protocols to ensure equal scan times for all sequences. This ensured that magnetization recovery was similar and supported a fair comparison of signal levels. On the scanner, the total scan time of all sequences were set to the sequence with the longest RF pulse duration, which was the LIBRE pulse with RF duration of 2.6 ms. The scanner automatically adds a dead time each 50 TRs, which corresponded to approximately 105 ms for FS, 45 ms for WE (1-1), 0 ms for LIBRE (2.6 ms), 20 ms for LIBRE (2.2 ms), 40 ms for LIBRE (1.8 ms) and 60 ms for LIBRE (1.4 ms).

*MRI volunteer study*

For the human studies, we had permission from the Institutional Review Board and written informed consent was obtained from all volunteers before the scan. Legs of healthy volunteers (n=7) were scanned to evaluate the LIBRE method in vivo and on a 3.0 T clinical MR scanner.

Identical LIBRE pulse parameters as those described above were tested and evaluated for fat suppression using a field of view (FOV) of $156 \times 192$ x 160 mm$^3$, a matrix size of $156 \times 192 \times 160$ (resulting in a voxel size of 1.0 x 1.0 x 1.0 mm$^3$), 50 k-space lines per segment using a linear reordering approach,  with  RF and gradient spoiling, RF excitation angle $= 10°$, echo times (TEs)  and repetition times (TRs) are dependent on pulse parameters and indicated in Table 1. Similar to the phantom experiments, additional MRI scans were performed using a standard acquisition, FS and WE (1-1).

*MRI data reconstruction and analysis*

Image reconstruction was performed by using the sum-of-squares of all channels. Following the MRI experiments, images were analyzed using ImageJ (National Institutes of Health, Bethesda, Maryland, USA). The level of fat suppression was determined by computing the average signal from ROIs drawn in compartments containing baby oil in the case of phantom experiments or subcutaneous lipids in the case of the in vivo experiments. Similarly, the average signals from muscle tissue (in vivo experiments) and agar (phantom experiments) were calculated, as well as the signals from the background noise. Subsequently, the signal-to-noise ratio (SNR) was computed as the ratio of the average signal and the standard deviation of the background noise. SAR values were recorded for all MRI protocols.

*Statistical methods*

Statistics were computed via two-tailed student t-tests for paired data with equal variance, corrected for multiple comparisons with Bonferroni. All data were expressed as mean ± standard deviation of the mean (SD). P<0.05 was considered statistically significant.

## RESULTS

*Numerical simulations*

Numerical simulations of the LIBRE pulse show the effect of magnetic field inhomogeneities, pulse duration, and excitation angle as a function of the pulse offset frequency on the magnetization magnitude as well as the phase (Fig. 2 and Supporting Fig. S1). These same effects were shown for the BORR pulse numerical simulation (Fig. 3). From these simulations, it can be inferred that there exists a range of LIBRE pulse properties that result in both optimal fat suppression and water excitation. It was demonstrated that the LIBRE pulse is insensitive to a large range of $\Delta B_0$ (Fig. 2g-i), which, conversely, translates to a fairly large stop band for lipids. These simulations also suggest that there exists a relatively large range of RF excitation angles for which water excitation and fat suppression are effective simultaneously. The signal response of water varies slowly as a function of the changing excitation angle. Since variations of the excitation angle may also be regarded as changes in $B_1$, the simulation results predict that the LIBRE pulse maintains both effective water excitation and concomitant fat suppression over a broad range of RF field inhomogeneities $\Delta B_1$. In general, the strategy for choosing experimental pulse parameters is to select an optimal combination of RF pulse frequency offset and pulse duration for fat suppression (Fig. 2b), and then to optimize the RF excitation angle for water excitation. Based on these simulations, four combinations of pulse parameters were chosen for further experimental validation with total pulse durations of 1.4, 1.8, 2.2 and 2.6 ms. The use of a LIBRE defined phase offset (Eq. 1) show maximized water excitation as well as fat suppression for the four abovementioned parameter combinations (Fig. 4).

*Experimental results*

The findings from the numerical simulations (Fig. 2) that predict optimal pulse settings corroborate well with the experimentally measured signals (Fig. 5). With LIBRE, the signal from the oil compartments is very well suppressed which leads to an almost indiscernible contrast between fat and the background signal irrespective of the LIBRE pulse duration. Simultaneously, the signal from agar remains largely unchanged when comparing the standard acquisition with those obtained with the four

LIBRE parameter settings (Fig. 5g and 5h). This leads to a consistently high agar SNR, and even applies for the shortest pulse duration.

While it is evident that all the investigated fat suppression techniques attenuate fat signal as compared to the non fat suppressed ("Standard") image acquisition (Fig. 6a-d), in the phantom data, it can be observed that it is most effective using LIBRE (Fig. 6a-d). Note that window and level are identical in all images. The related quantitative analysis in the same phantom shows statistically significant differences in fat SNR (Fig. 6e) comparing LIBRE with FS and WE (both p<0.004 and p<0.003 respectively). A SNR reduction of 47%, 81%, and 90% respectively was measured (Table 1) for FS, WE and LIBRE, meaning that there is almost no remaining fat signal using LIBRE. The SNR of the agar compartment was similar for the four different investigated MRI sequences (Fig. 6e).

The relative insensitivity to magnetic field inhomogeneities that were expected based on the simulations are clearly visible in the experimental results when looking at the phantom borders (green arrows, Fig. 6c). In these regions magnetic field inhomogeneities are most pronounced due to phantom-air interfaces, and this effect is clearly visible in the case of WE (1-1) (Fig. 6 c).

In the human in vivo experiments, using a frequency as optimized in the simulations, the findings from both theoretical predictions (Fig. 2) and experimental phantom results (Fig. 5 and 6) were corroborated (Fig. 7).

In vivo, and similar to what was found in the phantom experiments, all the tested approaches attenuated fat signal to various degrees when compared to the standard MRI sequence (Fig. 7). A more detailed account of the in vivo results are provided in the Supporting Information (Supporting Fig. S2). For improved clarity only images that were acquired with a LIBRE pulse duration of 2.2 ms and 1.4 ms are shown in this figure. The effectiveness of the fat suppression using LIBRE can be appreciated (Fig. 7d) when compared to FS (Fig. 7b) and WE (Fig. 7c). The lower efficiency of fat suppression with WE and FS is also apparent in areas close to air-tissue interfaces, where magnetic field inhomogeneities are typically increased (yellow arrows, Fig. 7). When compared to the quantitative measurements in the standard acquisition, the SNR of fat (Fig. 7e, Table 2) shows a decrease from $84.9 \pm 7.6$ to $33.5 \pm 2.8$ (FS), $31.0 \pm 3.5$ (WE), and $17.9 \pm 2.9$ for LIBRE (1.4 ms). LIBRE shows a significant decrease in the fat SNR compared with FS and WE (both p=0.00005). Consistent with numerical simulations and the in vitro findings, no significant changes in the SNR in muscle was observed (all p=NS) albeit a trend for a slightly reduced SNR was measured for the shortest investigated version of the LIBRE pulse.

## DISCUSSION

In this study we have developed, implemented, tested and characterized a flexible off-resonant RF water excitation pulse that leads to effective fat signal suppression without the need of a pre-pulse. It was shown that it is robust to magnetic field inhomogeneities, supports short RF pulse durations and remains effective at high magnetic field strength. The flexible phase reset $\varphi_2$ that is determined by both pulse duration and frequency (Eq. 1) provides a maximum excitation of the water magnetization in this type of pulse (Fig. 4).

This method was developed and implemented at 3 T, but the framework may also be applied to other magnetic fields strengths. It needs to be considered that the theoretical pulse duration scales inversely proportional to the Larmor frequency. Therefore, and if short pulse durations and short TR are required, these pulses may be more beneficial at higher magnetic field strength. LIBRE pulses show excellent fat suppression that seems robust against magnetic field inhomogeneities. Even though a deliberate optimization of the frequency offset and the pulse duration is mandatory, fat suppression remains effective over a relatively large range of parameter settings and the results from the numerical simulations were corroborated by both in vitro and in vivo findings. Since the level of water excitation and fat suppression depends on many pulse parameters that are inter-related, numerical simulations are required to make informed decisions on the optimal parameter setting when changing magnetic field strength or pulse duration. As the results from the in vitro and in vivo experiments were consistent with the behavior predicted by the numerical simulations, we can also deduce that these simulations provide meaningful guidance for optimal RF pulse parameter definition.

Although the LIBRE pulses are relatively long (1-2 ms) and inevitably result in increased TR and TE when compared to regular broadband excitation pulses (0.1-0.5 ms) preceded by a fat saturation pre-pulse, the LIBRE pulses are low in RF energy deposit (or SAR), as this scales with the RF pulse amplitude squared. When applying an equal TR and scan time for each MRI method, the SAR deposition in the LIBRE method compared to FS and WE was decreased (Fig. 8). In some implementations of the conventional 1-1 WE pulse the sub-pulse durations may be changed to further reduce SAR. In many applications and particularly at higher field, the imaging sequences are limited by SAR for safety reasons. Therefore, the total imaging duration may not necessarily suffer from the longer TR that is inherent to the LIBRE pulses. The total RF pulse duration with LIBRE is shorter but does not differ significantly from that of a binomial 1-1 WE pulse (1.8 ms on our system) that was available to us at 3 T. However, when compared to WE, and consistent with the lower SAR mentioned above, LIBRE allows for higher RF excitation angles to be used, which may be a distinct advantage for some applications. A comparison with more advanced spectral spatial pulses (14) was not performed in this study but may be of interest for a complete evaluation. In addition to the lower SAR, simulations also predict that $M_{xy}$ remains relatively constant over a broad range of RF excitation angles (Fig. 2f), suggesting that the LIBRE method may have a low sensitivity to $B_1$ inhomogeneities. These simulations also predict that the LIBRE method has a lower sensitivity to $B_1$ inhomogeneities compared with the BORR method (Fig. 2f and Fig. 3f). The numerical results of both methods (Fig. 2 and Fig. 3) also suggest an increased fat suppression bandwidth using LIBRE (Fig. 2h and Fig. 3h). The phase reset of $\varphi_2$ that depends on the pulse duration and frequency (Eq. 1) is an important difference compared with the BORR method, where a fixed phase shift of $\pi$ is applied (Fig. 4). In the latter case, efficient fat suppression is obtained, but water excitation may not be optimal, while the LIBRE method simultaneously provides both effective fat suppression and water excitation (Fig. 4).

A typical binomial 1-1 WE pulse consist of two short pulses with a inter-pulse delay corresponding to a 180° phase evolution between water and fat protons (14,17-19), which is 1.2 ms at 3 T. This is important in order to rotate the fat magnetization back to alignment with $B_0$ without transverse component, while keeping the water magnetization transverse. It is exactly this feature of the WE pulse that makes it so sensitive to magnetic field inhomogeneities, as the frequency changes caused by $\Delta B_0$ result in a phase evolution different from 180°. Especially at higher magnetic field strengths, $\Delta B_0$ is bound to increase, and these magnetic field inhomogeneities are spatially dependent and lead to variable Larmor frequencies across the field of view. Related effects can especially be observed at air-tissue interfaces where more conventional fat suppression techniques may more easily break down (regions indicated by yellow arrows Fig. 7). It is therefore desirable that fat suppression methods are as insensitive as possible to such inhomogeneities to be most effective. The $M_{xy}$ signal response for the LIBRE pulse also has a low sensitivity to small variations in $\Delta B_0$ for a RF frequency offset f in the range from 100 to 300 Hz (Fig. 2g, h, and i). While higher order binomial water excitation pulses, such as a 1-2-1 or 1-3-3-1 may achieve broader suppression bandwidths compared to 1-1 and thus perform better, this comes at the expense of a longer pulse duration and TR. Because of such prolonged pulse durations, these higher order binomial pulses were not considered in this study. However, for applications where an increase in TR is not considered a limitation, higher order binomial pulses may perform equally well or better than LIBRE. While this remains to be studied, such binomial pulses afford the advantage of spatial selectivity when compared to LIBRE.

In this study, we also compared images acquired with the LIBRE strategy to those obtained with CHESS for fat signal suppression. For CHESS we used a relatively lengthy acquisition window (~200 ms for FS), which is a disadvantage as fat magnetization recovers during the acquisition period. While the performance of CHESS would benefit from a shorter acquisition window, this inevitably increases scan time as well as the RF energy deposit. It has to be noted that fat suppression with CHESS can be further improved using a center-out instead of a linear profile order. Cartesian sampling schemes were also considered, which may be more forgiving in terms of the effectiveness of fat saturation using a pre-pulse. Particularly for radial imaging where every single profile traverses the center of k-space, fat saturation using pre-pulses may be less effective. T1 based fat suppression approaches where fat is either fully or partially inverted were not investigated in this study. While these methods have shown to work very effectively (6-8), the time needed for signal nulling may adversely affect overall scanning time for certain applications. LIBRE provides only fat suppression and not water fat separation and it has a relatively long RF pulse duration. Therefore DIXON or IDEAL techniques may provide valuable alternatives albeit at the expense of scanning time and motion sensitivity on the one hand, but which can easily be combined with slice selective excitation on the other hand.

In fact, and as already mentioned in the context of binomial pulses above, a significant drawback of our method includes that the LIBRE pulses are not spatially selective. Therefore, it may not easily be applicable to 2D methods or 3D approaches in which reduced volumetric coverage is desired. The property of the LIBRE pulse to flexibly adjust some of its parameters , with effective total RF durations as low as 1.4 ms, emphasizes an important difference with the BORR implementation with a total RF duration of 2.6 ms and 3.2 ms (21,22). In this study we have not explored RF pulse durations below 1.4 ms, but we foresee that this may be possible. In 3D brain applications such as functional MRI at high magnetic field using EPI (16,27), the use of CHESS based fat suppression leads to a reduced imaging efficiency and increased RF power deposition. To circumvent this, the use of long on-resonant rectangular excitation pulses (~2.5 ms at 3T) were investigated to achieve simultaneous fat suppression, scan time, as well as SAR reduction in 3D EPI (16). The use of LIBRE may offer considerable advantages here because it may further reduce the RF duration, while still maintaining effective fat suppression.

Currently, the LIBRE pulse parameters can manually be selected from the scanner user interface (UI) and in our study, the choice of these parameters was informed by the simulation results. While such an approach may not be practical in daily clinical routine, a range of pulses optimized for certain applications could be stored on the system and selected from the user interface for improved ease of use.

The main consideration for improving the ease of use are the desired TR and possible SAR constraints, therefore, several LIBRE pulses of fixed duration may be chosen from the UI. Based on this user input, all other pulse parameters may then be set automatically by the system. Changing the RF excitation angle will not affect the LIBRE pulse choice. So far we have investigated the use of LIBRE in 3D Cartesian MRI only. However, it is now also planned to optimize LIBRE for its use in 3D radial imaging (28,29) that is inherently sensitive to incomplete fat suppression at higher field strengths as mentioned above.

## CONCLUSION

A water excitation method that homogeneously and effectively suppresses fat signal has been developed, implemented, characterized, and successfully tested in vitro and in vivo at 3T. The LIBRE method was shown to be robust to magnetic field inhomogeneities and effective at short RF pulse durations.

**Acknowledgements**

R'Equip SNF grant 326030_150828, NANOTERA

**Table 1**

Scan parameters for the 3D gradient echo MRI pulse sequences used in this study.

| Parameter | Scan parameters for different sequences | | | | | | |
|---|---|---|---|---|---|---|---|
| | Standard | FS | WE (1-1) | LIBRE (2.6 ms) | LIBRE (2.2 ms) | LIBRE (1.8 ms) | LIBRE (1.4 ms) |
| RF duration (total in ms) | 0.50 | 0.50 | 1.73[a] | 2.60 | 2.20 | 1.80 | 1.40 |
| TE (ms) | 1.80 | 1.80 | 2.42 | 2.86 | 2.66 | 2.46 | 2.26 |
| TR (ms) | 4.0 | 4.0 | 5.2 | 6.1 | 5.7 | 5.3 | 4.9 |
| RF Pulse offset (Hz) | 0 | 0 | 0 | 300 | 500 | 700 | 1000 |

[a] The WE (1-1) pulse consists of two RF sub-pulses each with a duration of 0.50 ms.

**Table 2**

In phantoms in vitro (n=5), the signal-to-noise ratio was measured in the agar and in the oil compartment. In healthy volunteers in vivo (n=7), the same analyses were conducted for muscle and fat tissue. Percent fat SNR reduction and statistical significance relative to no fat suppression (Standard) is provided in parentheses.

| | | Signal-to-noise ratios of different MRI methods | | | | | | |
|---|---|---|---|---|---|---|---|---|
| | | Standard | FS | WE (1-1) | LIBRE (2.6 ms) | LIBRE (2.2 ms) | LIBRE (1.8 ms) | LIBRE (1.4 ms) |
| Phantom (n=5) | Oil | 90.6 ± 20.4 | 48.0 ± 13.0 (-47%, | 17.4 ± 3.0 (-81%, | 9.1 ± 1.3 (-90%, | 9.5 ± 1.2 (-90%, | 9.2 ± 1.2 (-90%, | 9.3 ± 1.2 (-90%, |
| | Agar | 59.9 ± 8.1 | 57.7 ± 7.1 | 56.8 ± 8.6 | 59.3 ± 9.9 | 64.5 ± 9.5 | 60.6 ± 7.5 | 56.9 ± 6.4 |
| Volunteer (n=7) | Lipid tissue | 84.9 ± 7.6 | 33.5 ± 2.8 (-57%, p=0.000006) | 31.0 ± 3.5 (-60%, p=0.000002) | 16.3 ± 2.2 (-79%, p=0.00008) | 16.9 ± 1.9 (-78%, p=0.000006) | 18.1 ± 1.0 (-77%, p=0.000054) | 17.9 ± 2.9 (-77%, p=0.000007) |
| | Muscle tissue | 62.6 ± 6.6 | 59.2 ± 6.1 | 62.5 ± 6.7 | 59.2 ± 6.2 | 58.9 ± 6.5 | 59.25 ± 8.9 | 54.8 ± 6.7 |

**Figure Captions**

**Figure 1**

Pulse sequence diagram of a Cartesian 3D GRE sequence with implementation of LIBRE RF excitation pulses. As described in the main text, the LIBRE pulse consists of two rectangular sub-pulses. Each rectangular sub-pulse has the same offset frequency f, duration τ, RF excitation angle α, yet an unequal phase φ. Here, the phase offset of the first RF subpulse is zero ($\varphi_1=0$), while the relative phase offset of the second RF sub-pulse is determined by the frequency and duration of the first RF sub-pulse, given by $\varphi_2 = \varphi_1 + 2\pi f\tau$.

**Figure 2**

Numerical simulation of the transverse magnetization $M_{xy}$ of water (left column), fat (middle column) and their respective difference (right column), immediately after RF excitation with the LIBRE pulse. Results are shown as function of a range of different pulse parameters at 3 T. (a-c) Simulations of $M_{xy}$ as function of the RF sub-pulse duration τ and of the relative frequency offset f.  (d-f) $M_{xy}$ as function of the RF excitation angle α and f. (g-i) $M_{xy}$ as function of the magnetic field inhomogeneities $\gamma\Delta B_0$ and f. Asterisks indicate the LIBRE parameters that were evaluated in phantom and volunteer experiments. The white dotted line indicates the LIBRE parameters that result in a phase difference of π between both sub-pulses. The corresponding phases of the transverse magnetization $M_{xy}$ are represented in Supporting Figure S1.

**Figure 3**

Numerical simulation of the amplitude of the transverse magnetization $M_{xy}$ of water (left column), fat (middle column) and their respective difference (right column), immediately after RF excitation with the BORR pulse. Results are shown as function of a range of different pulse parameters at 3 T. (a-c) Simulations of $M_{xy}$ as function of the RF sub-pulse duration τ and of the relative frequency offset f (with $\gamma\Delta B_0=0$ and α=10°) .  (d-f) $M_{xy}$ as function of the RF excitation angle α and f (with $\gamma\Delta B_0=0$ and τ =1.1 ms). (g-i) $M_{xy}$ as function of the magnetic field inhomogeneities $\gamma\Delta B_0$ and f (with τ=1.1 ms and α=10°).

**Figure 4**

Numerical simulation of the transverse magnetization $M_{xy}$ after a single RF excitation with two off-resonant sub-pulses where the relative phase offset of the second sub pulse $\varphi_2$ was varied over a range from 0 to 2π (360°). $M_{xy}$ was then computed as function of the tissue frequency and phase offset of the second subpulse ($\varphi_2$).  Four different RF pulse settings were simulated that were also used in the phantom and volunteer experiments: a) Sub-pulse duration of 1.3 ms and RF frequency of 300 Hz, b) 1.1 ms and 500 Hz, c) 0.9 ms and 700 Hz, d) 0.7 ms and 1000 Hz. It can be seen that maximum water excitation is achieved when $\varphi_2$ corresponds to a LIBRE defined phase offset according to Eq. 1 with $\varphi_2=2\pi ft$ (horizontal dashed black lines), as it lies at the intersection of the water frequency (0 Hz, vertical dashed lines) and the maximum observed $M_{xy}$ at that frequency ("circular" dashed lines). The relative phase offset $\varphi_2$ corresponding to $\pi$ (grey dashed line) is also depicted. A large fat suppression bandwidth (grey arrows) can be observed.

**Figure 5**

Images and results obtained in a cylindrical phantom using five different MRI methods at 3 T. (a) Schematic of the cylindrical phantom with indications of the agar and oil compartments. (b-f) The images represent a reformat of the phantom within a 3D volume acquired using different 3D GRE acquisitions. (b) MR image obtained without the use of fat suppression, (c-f) images obtained using four different parameter sets for the LIBRE water excitation as indicated in the figures. (g) SNR for oil (h) and agar relative to the background signal (b-f). Image intensities were scaled for identical window and level settings.

**Figure 6**

Images and results obtained in a cylindrical phantom using four different MRI methods at 3 T. The images represent two orthogonal reformats from a 3D volume acquired using a 3D GRE acquisition. (a) MR images obtained with the standard MRI sequence, (b) images obtained using conventional fat saturation (FS), (c) images obtained using conventional 1-1 binomial water excitation (WE), (d) image obtained using LIBRE pulses with a duration of 1.4 ms. (e) SNR computed in both the oil and agar compartments for all four sequences. Significant changes (**p<0.005) were observed using LIBRE fat suppression compared with FS and WE. Image intensities were scaled to identical window and level settings. Arrows indicate regions of inhomogeneities, either due to air-phantom interfaces (green arrows) or compartment border (yellow arrows) that hinder fat suppression in WE.

**Figure 7**

Images obtained using five different MRI methods performed in a human volunteer (knee) at 3 T. (a) MR images obtained without the use of fat suppression, (b) images obtained using conventional fat saturation (FS), (c) images obtained using conventional 1-1 binomial excitation (WE), (d) images obtained using LIBRE pulses of 1.4 ms duration, respectively. Note the enhanced and uniform fat suppression of subcutaneous fat using the LIBRE pulses. Arrows indicate regions within the knee that show incomplete fat suppression most likely due to magnetic field inhomogeneities. All image intensities were scaled to identical window and level settings. (e) Comparison of SNR computed in different regions (fat and muscle) of legs of human volunteers obtained with different fat saturation approaches including LIBRE with four different parameter settings. Significance levels are indicated for the comparison of LIBRE (1.4 ms) fat suppression with conventional FS and WE.

**Figure 8**

Comparison of SAR values recorded during MRI experiments with equal TR and scanning time.

**Supporting Figure Captions**

**Supporting Figure S1**

Numerical simulation of the phase of the magnetization of water (left column), and fat (right column) immediately after RF excitation with the LIBRE pulse. Results are shown as function of a range of different pulse parameters at 3T. (a-b) Simulations of the phase as function of the RF sub-pulse duration $\tau$ and of the relative frequency offset f.  (c-d) Phases as function of the RF excitation angle $\alpha$ and f. (e-f) Phase as function of the magnetic field inhomogeneities $\gamma \Delta B_0$ and f. The corresponding magnitudes of the transverse magnetization $M_{xy}$ are represented in Figure 2 of the manuscript.

**Supporting Figure S2**

Images obtained using five different MRI methods performed in a human volunteer (knee) at 3 T. (a) MR images obtained without the use of fat suppression, (b) images obtained using conventional fat saturation (FS), (c) images obtained using conventional 1-1 binomial excitation (WE), (d and e) images obtained using LIBRE pulses of 2.2 and 1.4 ms duration, respectively. Note the enhanced and uniform fat suppression of subcutaneous fat using the LIBRE pulses. Arrows indicate regions within the knee that show incomplete fat suppression most likely due to magnetic field inhomogeneities. All image intensities were scaled to identical window and level settings.

Figure 1

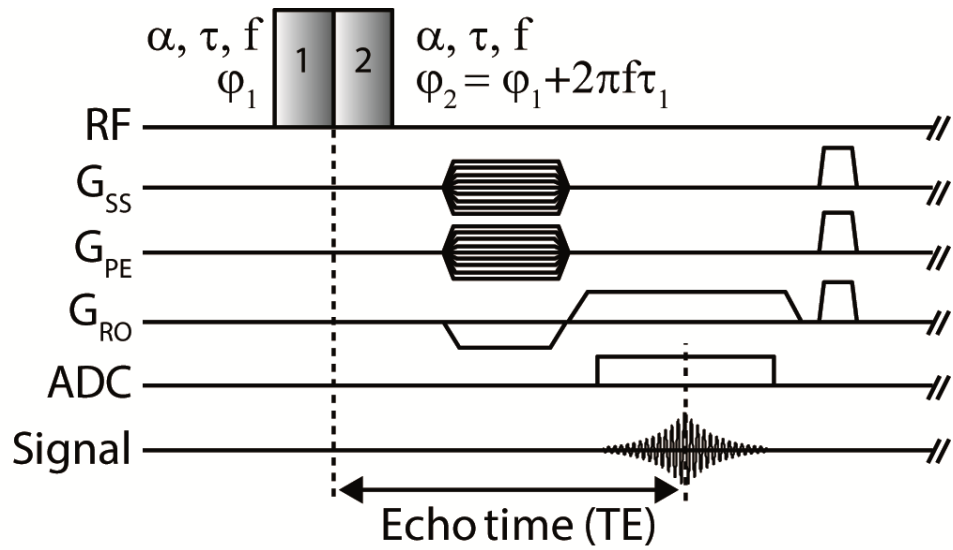

Figure 2

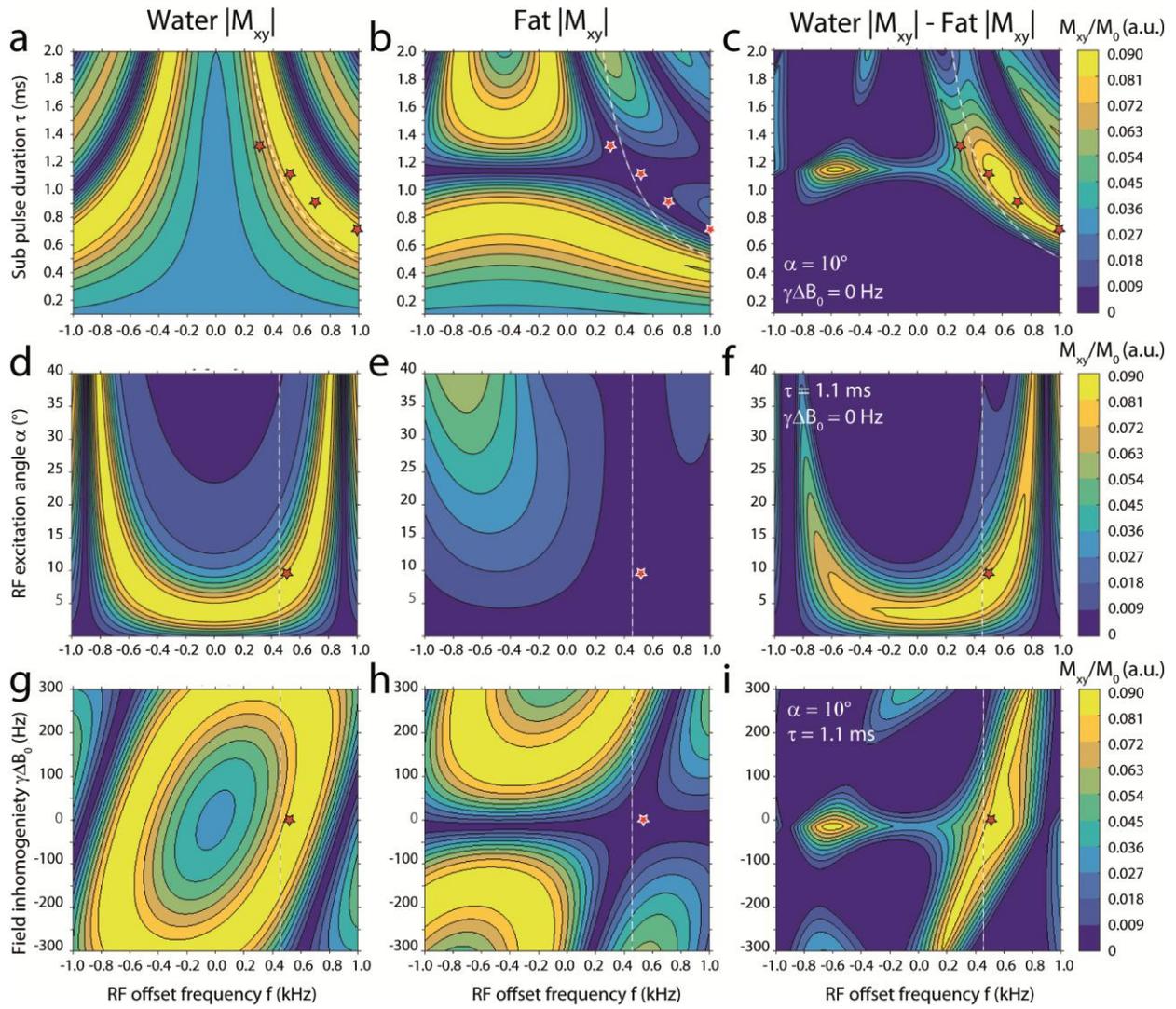

Figure 3

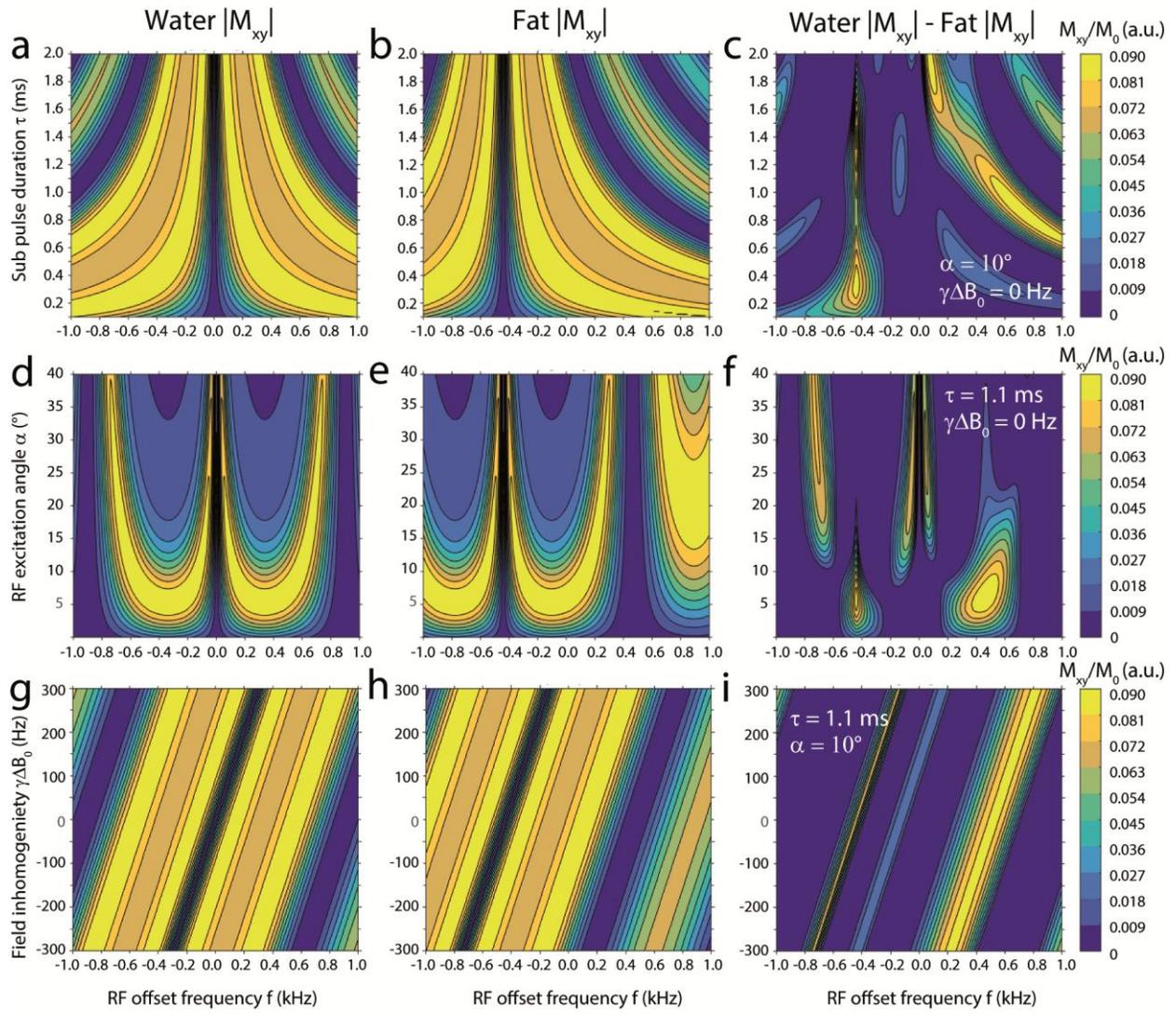

Figure 4

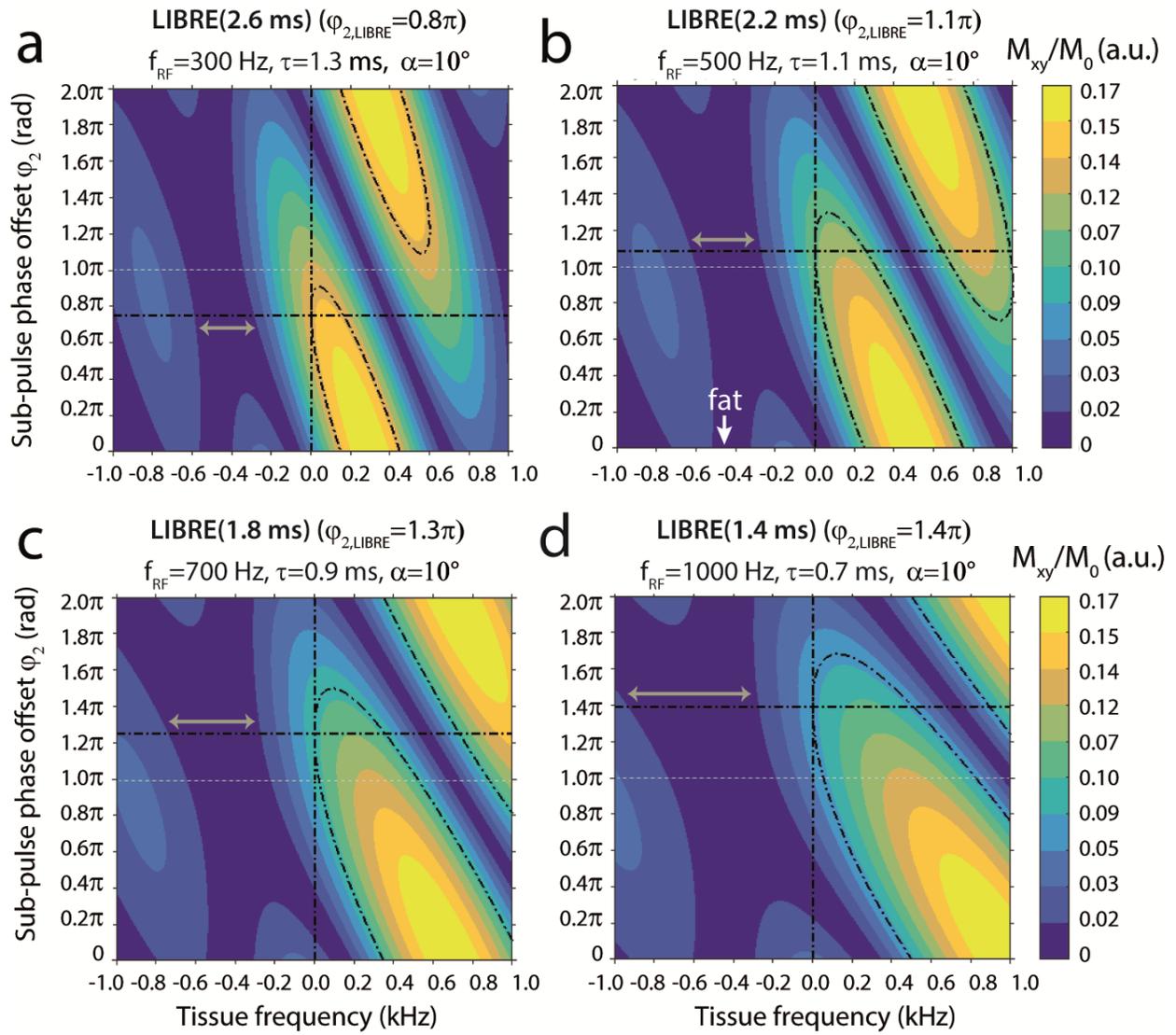

Figure 5

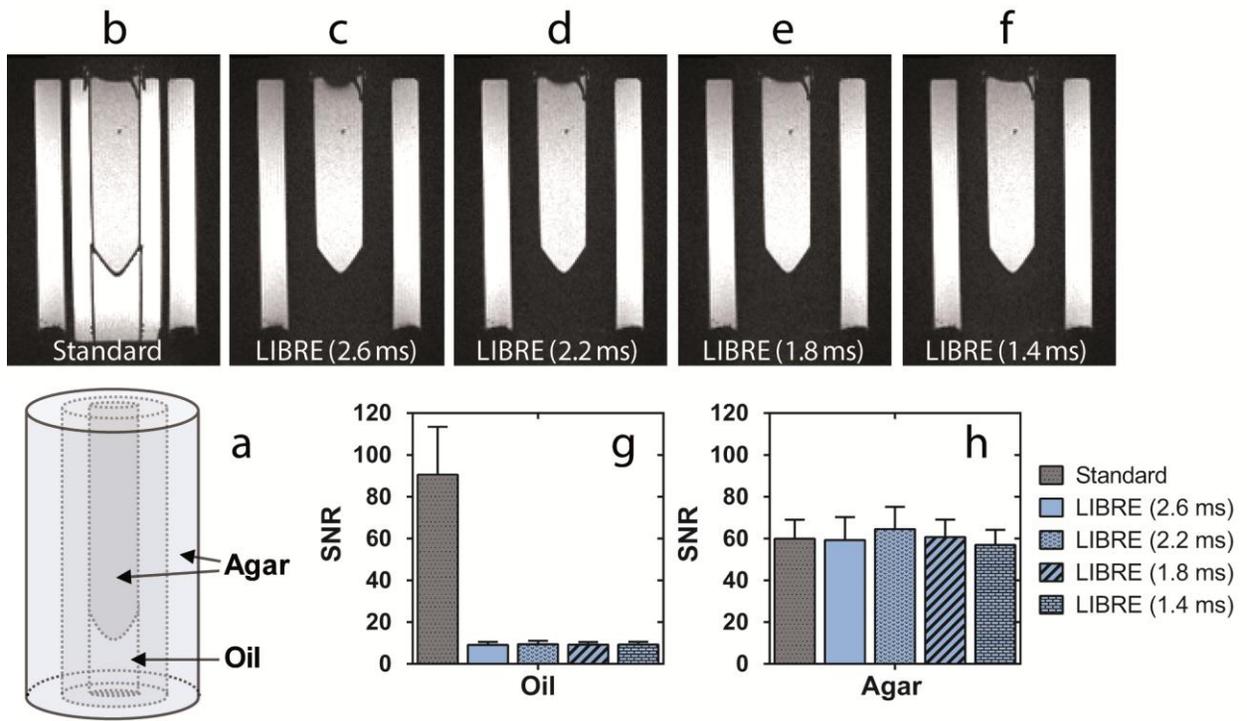

Figure 6

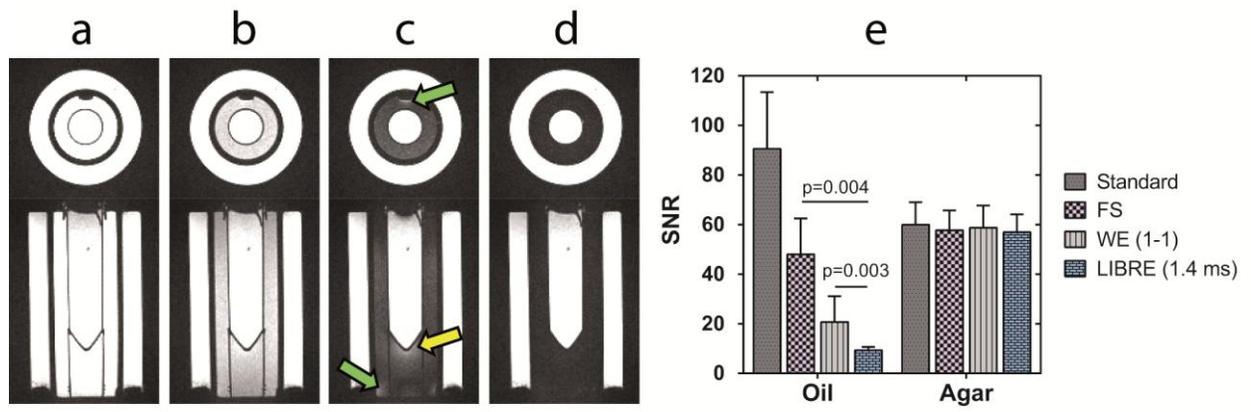

Figure 7

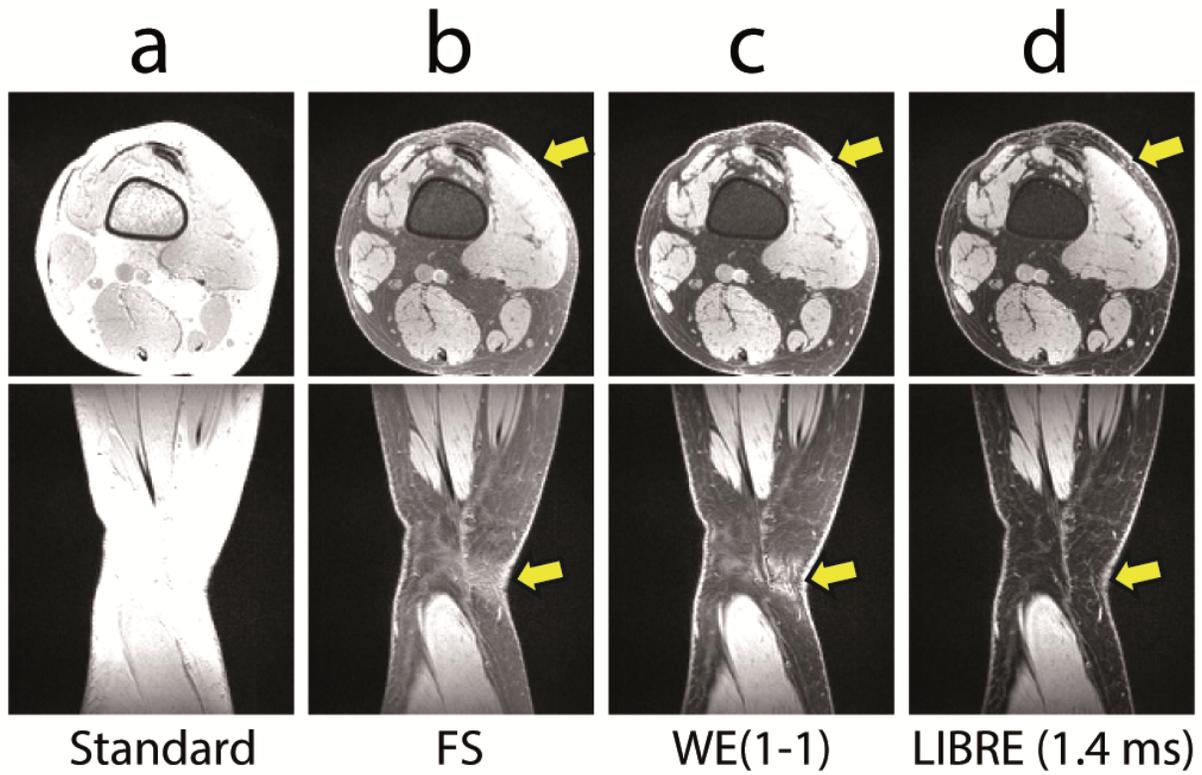

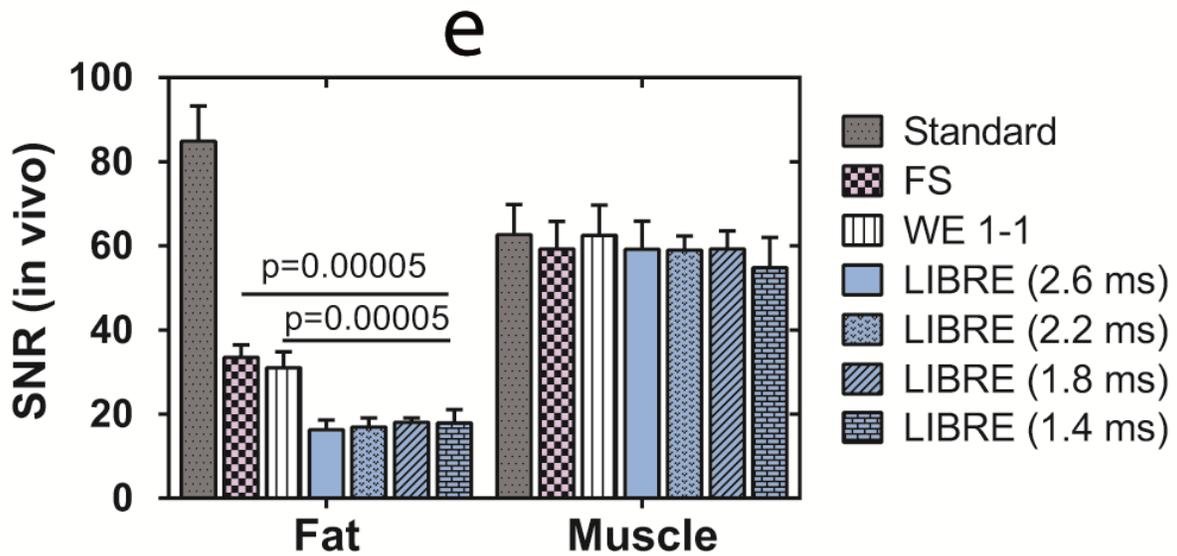

Figure 8

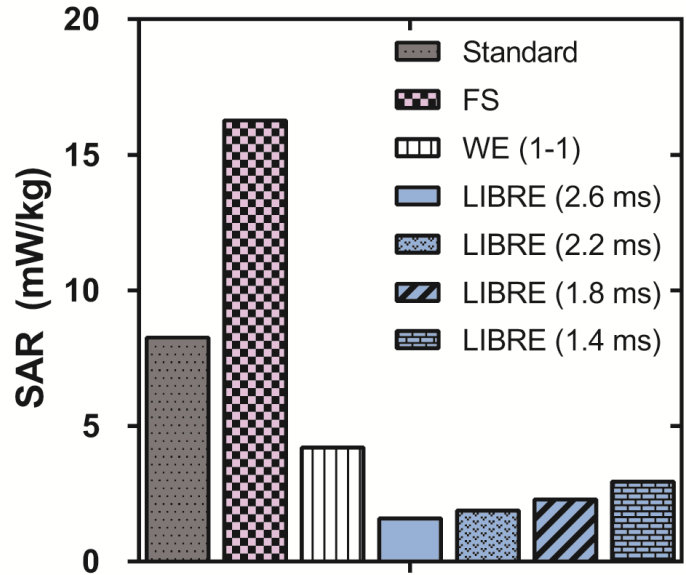

Supporting Figure 1

Water arg(M_xy)    Fat arg(M_xy)

Supporting Figure 2

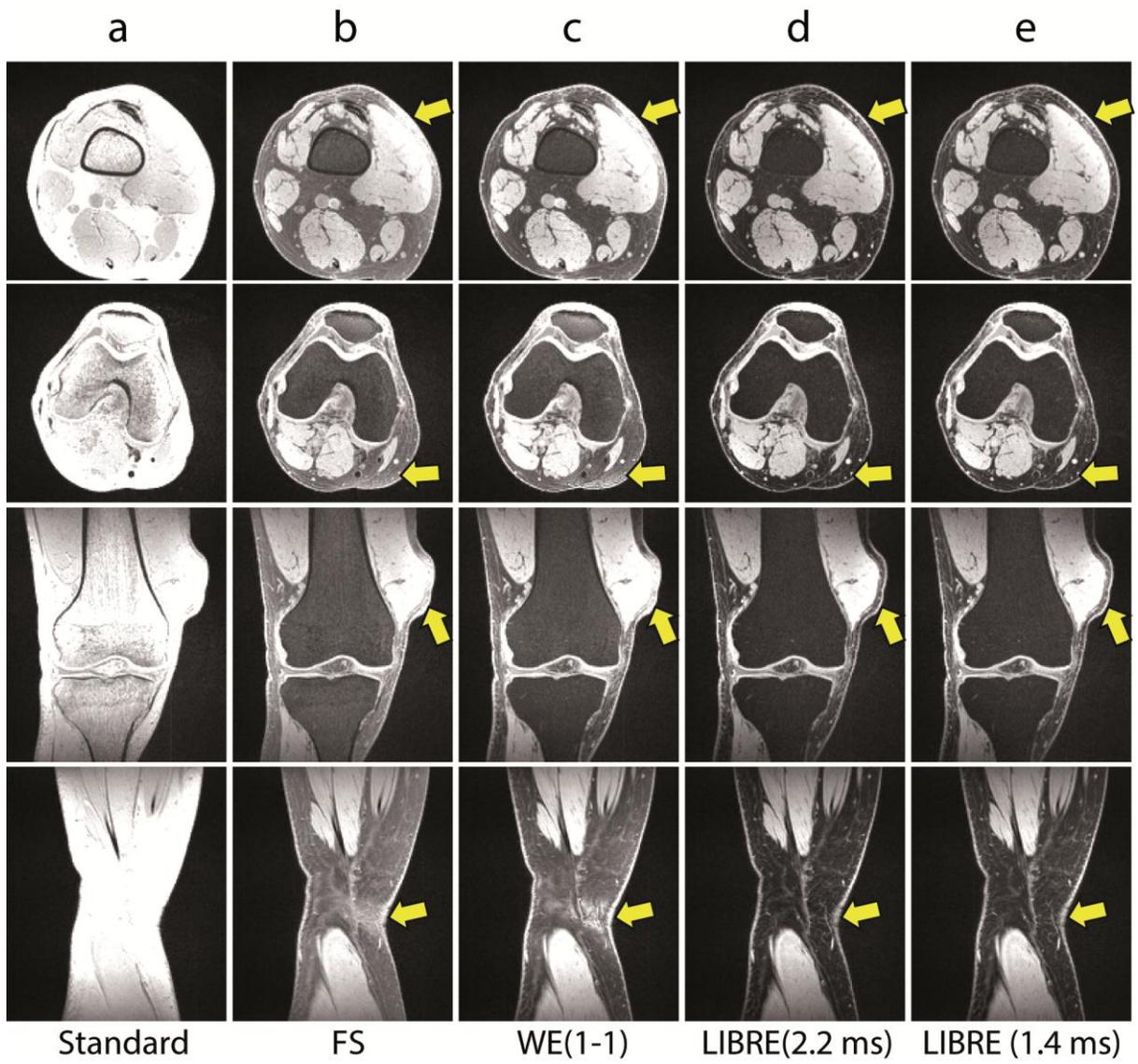